\documentclass[sigconf,nonacm]{acmart}

\usepackage{amsmath}
\usepackage{booktabs}
\usepackage{graphicx}
\usepackage{microtype}
\usepackage{multirow}
\usepackage{xcolor}

\setcopyright{none}
\renewcommand\footnotetextcopyrightpermission[1]{}
\title{Which Reranking Conclusions Survive the Answer Interface?\texorpdfstring{\\}{ }A Prospective Finite-Orbit Audit}

\author{Wenzhang Du}
\affiliation{%
  \institution{Independent Researcher}
  \country{China}}
\email{dqswordman@gmail.com}

\begin{document}

\begin{abstract}
Rerankers are increasingly evaluated through downstream language-model answers.  This raises a retrieval-measurement question: if only the reader's answer interface changes, should we reach the same conclusion about BM25 versus BGE-v2-m3?  We prospectively audit their claim-paired effect on RAGuard and FEVER with four readers.  Retrieval policies, evidence, claims, and context depth remain fixed while semantic-to-label binding, A/B versus X/Y vocabulary, and option order form eight task-equivalent interfaces.  We ask whether the estimated retrieval-policy effect, its ordering, or selection value changes.  None of the six confirmatory settings showed statistically certified interface variation above the prespecified 0.015 materiality threshold, and none showed a certified reversal of the BM25--BGE ordering.  Selector disagreement reaches 33.5\% in one environment, yet none of eight environments establishes the prespecified material held-out value difference.  These results do not support broad replicated instability, but they do not prove universal invariance: five settings remain too uncertain to satisfy the prespecified higher-order equivalence condition.  They show why retrieval evaluations should separate policy level, interface stability, policy ordering, and selection value.  When stability is unverified, a uniform average over the enumerated interfaces with explicit variation bounds avoids privileging one interface.
\end{abstract}



\maketitle

\section{Introduction}

When a language-model reader is used to evaluate a reranker, what exactly counts as a conclusion about retrieval?  Consider a fact-checking claim evaluated with evidence from BM25 or from a BGE reranker.  If the reader is more accurate with BGE evidence, we may conclude that reranking helps.  That conclusion is about a retrieval policy, even though it is observed through a downstream answer.

The answer interface introduces an arbitrary coordinate into this measurement.  The semantic task does not change when SUPPORTS is encoded as A instead of X, bound to the other label, or displayed second rather than first.  The evidence, claim, and retrieval policy are also unchanged.  A comparative retrieval claim should therefore not depend silently on these choices.  The question is not merely whether a reader is prompt-sensitive, but whether that sensitivity changes what we conclude about \emph{BM25 versus BGE}.

Prior work already establishes label and order sensitivity, performance variation across prompts, semantic reformulation effects, and sensitivity to RAG configuration~\cite{zheng2024pride,liusie2024general,polo2024prompteval,thomas2026forminv,hsia2025ragged}.  Prompt wording can even rival the ranking algorithm in zero-shot LLM rankers~\cite{sun2025promptrankers}.  We study a narrower unresolved question: how stable is a comparative retrieval-policy effect measured through a downstream reader?  This question separates four conclusions.  At the policy level, does BGE help on average and does its gain differ by verdict?  How much does that effect vary across equivalent interfaces?  Does the BM25--BGE ordering change?  And does an interface-specific estimate change which ranker we select or the held-out value of that choice?

We audit the complete orbit of three binary interface factors: semantic-to-label binding, label vocabulary (A/B or X/Y), and displayed option order.  For each claim, BM25 and BGE expose the same number of documents from the same candidate pool; only the answer rendering changes.  Four open readers are scored by teacher-forced complete-sequence likelihood on RAGuard and FEVER.  Six dataset--reader environments form the prospective confirmatory set, while two previously studied RAGuard checkpoints serve as descriptive bridges.

The prospective test does not support the broad replicated-instability hypothesis.  None of the six confirmatory environments meets the prespecified 0.015 material-instability rule after simultaneous correction, no interface pair certifies a reversal of the BM25--BGE ordering, and no selector test establishes a material held-out value difference.  This does not establish universal interface invariance: five environments remain too uncertain for the prespecified higher-order equivalence condition.  The result instead bounds what the motivating anomaly supports across datasets and reader families.

Our contributions are threefold:
\begin{itemize}
    \item We formulate a retrieval-policy evaluation problem that separates policy level (average gain $g$ and verdict differential $h$), interface stability, policy ordering, and downstream ranker-selection value.
    \item We conduct a two-dataset and four-reader audit, prespecified before downstream evaluation, over all eight combinations of three task-equivalent interface factors, using claim-paired simultaneous inference and a held-out selector test.
    \item We show that the motivating instability does not replicate as a broad material phenomenon under this design: 0/6 confirmatory environments meet the instability rule, no policy-sign reversal is certified, and selector disagreement does not yield a demonstrated material held-out value difference.
\end{itemize}

\begin{figure*}[t]
  \centering
  \includegraphics[width=\textwidth]{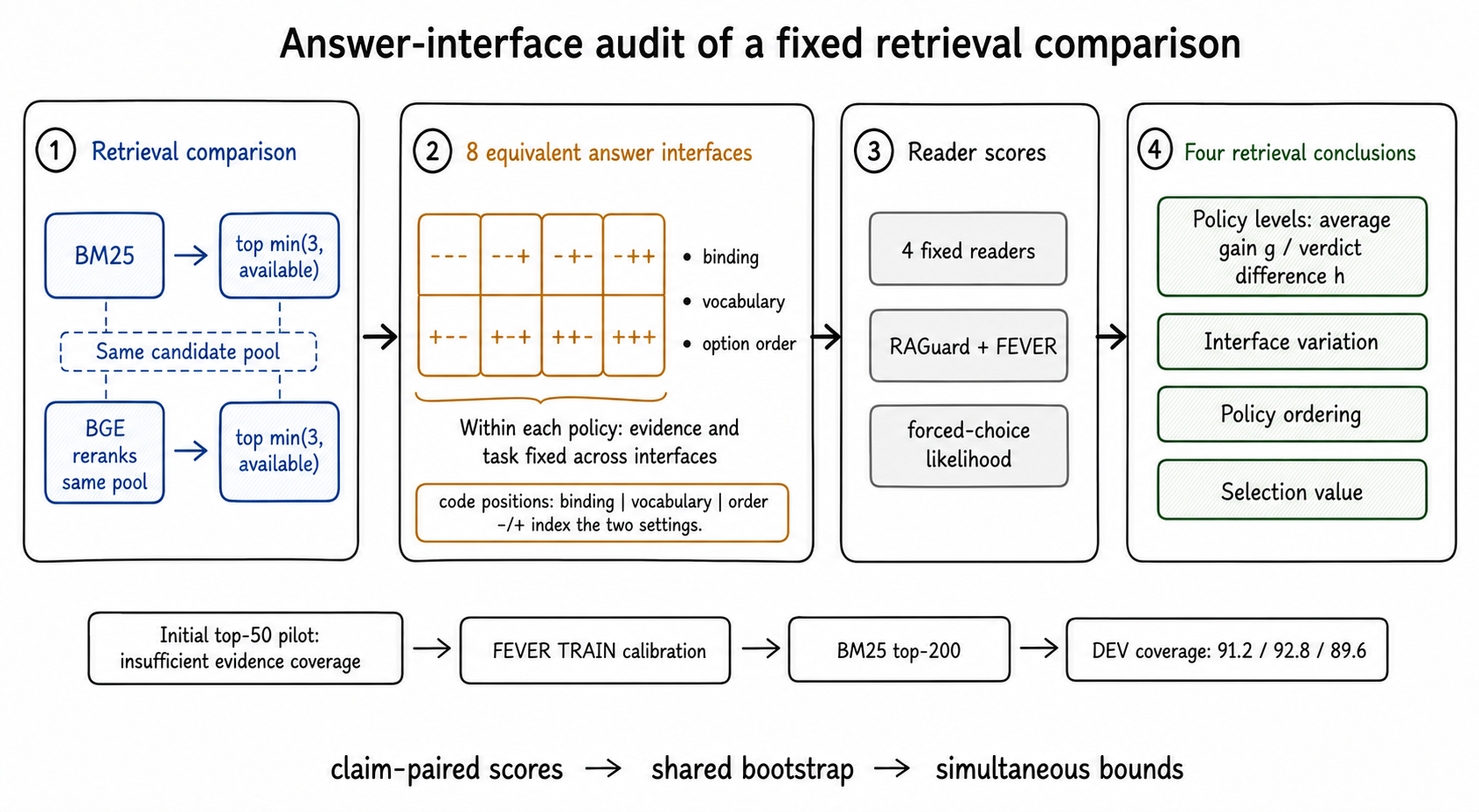}
  \caption{How we audit a fixed retrieval comparison.  BM25 and BGE use the same candidate pool and context depth; within each policy, evidence and task are fixed across interfaces.  Each three-sign code indexes binding, vocabulary, and option order.  We keep four retrieval questions separate: policy levels ($g$ and $h$), interface variation, policy ordering, and selection value.  The FEVER strip shows pre-evaluation calibration, not an outcome-dependent change.}
  \Description{A four-stage hand-drawn diagram. BM25 and BGE retrieve from the same candidate pool; eight equivalent answer interfaces vary binding, vocabulary, and option order, with a visible legend defining the code positions; four fixed readers score RAGuard and FEVER by forced-choice likelihood; and four output cards show policy levels, interface variation, policy ordering, and selection value. A smaller strip shows the FEVER training calibration and development-set evidence coverage.}
  \label{fig:overview}
\end{figure*}

\section{Related Work}

\subsection{RAG and retrieval evaluation}

RAG systems couple a retrieval policy to a language-model reader, so a measured end-to-end gain combines document access with the reader's use of that evidence~\cite{lewis2020rag}.  FEVER makes this coupling explicit through document retrieval, evidence selection, and verification~\cite{thorne2018fever}.  RAGuard then probes a difficult part of the same chain: how readers respond when naturally retrieved evidence supports or misleads the final verdict~\cite{zeng2025raguard}.  Our study uses these fact-checking settings because the semantic decision is binary while the retrieval policy remains a first-class experimental variable.

Several lines of work map sources of RAG variability.  RAGGED varies retrieval method, reader, context quantity, and prompt configuration to examine system-level stability~\cite{hsia2025ragged}.  Evidence-interface work distinguishes whether support is available from how retrieved support is organized for the reader~\cite{liao2026evidence}.  These studies ask how configurations or evidence presentations affect system performance.  We instead freeze one comparative treatment---BM25 versus BGE reranking over the same candidate pool and context depth---then vary only the downstream answer coordinates.  The resulting object is not reader accuracy alone, but the \emph{difference} in reader accuracy attributable to the retrieval policy.

Prompt sensitivity also appears inside retrieval models.  In zero-shot LLM ranking, prompt wording can alter effectiveness as much as the ranking method itself~\cite{sun2025promptrankers}.  That intervention changes the prompt used by the ranker.  Here BM25 and BGE outputs are held fixed within each paired comparison; the varied prompt component belongs only to the reader that measures their downstream effect.  This separation is essential to our scope: we do not claim to characterize all RAG instability, only whether an answer interface changes a conclusion about a fixed retrieval comparison.

\subsection{Interface and prompt robustness}

Multiple-choice studies show that answer labels are not neutral measurement devices.  PriDe estimates and removes permutation-dependent selection bias, and general permutation debiasing learns from permuted alternatives~\cite{zheng2024pride,liusie2024general}.  Other work shows that first-token probabilities can disagree with generated text answers, that parsing the text can improve robustness, and that token-based decision criteria can hide stronger semantic behavior~\cite{wang2024firsttoken,wang2024looktext,cho2025hidden}.  These results motivate our complete label scoring and our explicit diagnostic comparison with free generation.  They do not, by themselves, determine whether the \emph{difference between two retrieval policies} is stable.

PromptEval treats performance across prompt variants as a distribution rather than a single score~\cite{polo2024prompteval}.  FormInv studies measurement consistency and model-ranking changes under verified semantic reformulations, including regime-aware model selection~\cite{thomas2026forminv}.  Crossed symmetrization separates answer wording, order, and semantic attachments in binary judgments~\cite{huang2026yesno}, while permutation consensus applies repeated orderings to listwise factuality evaluation~\cite{huang2026pcfjudge}.  Our factorial orbit and uniform averaging draw on these established ideas; the Walsh representation is likewise an analysis instrument, not a new decomposition.

The closest prior work studies invariance of model scores or rankings under equivalent prompts or semantic forms.  Our setting adds a comparative retrieval layer: each interface yields a paired BM25-to-BGE effect on the same claims.  We ask whether its orbit-average level, verdict heterogeneity, variation and sign, and held-out policy-selection value lead to the same retrieval conclusion.  This prospective retrieval-specific evaluation question is our narrow contribution; we do not claim a new general principle of permutation robustness.

\section{Auditing Retrieval-Policy Conclusions}

\subsection{Average and verdict-dependent policy effects}

For each answer interface, we first measure how much BGE changes accuracy relative to BM25 separately on positive and negative claims.  Their average asks whether BGE helps overall.  Half their difference asks whether the gain depends on the claim's semantic class.

Formally, let $z=(z_1,z_2,z_3)\in\{-1,+1\}^3$ encode binding, vocabulary, and option order.  For claim $i$, ranker $r\in\{B,S\}$ (BM25 or semantic BGE), semantic state $y\in\{+,-\}$, and interface $z$, complete-sequence scoring yields a semantic decision $\widehat y_{irz}$.  Define the class-conditional paired accuracy difference
\begin{equation}
  \Delta_y(z)=\mathbb{E}_i\!\left[\mathbf{1}\{\widehat y_{iSz}=y\}-\mathbf{1}\{\widehat y_{iBz}=y\}\mid y_i=y\right].
\end{equation}
We analyze
\begin{equation}
  g(z)=\tfrac12\{\Delta_+(z)+\Delta_-(z)\},\qquad
  h(z)=\tfrac12\{\Delta_+(z)-\Delta_-(z)\}.
\end{equation}
Here $g$ is the balanced average policy effect and $h$ is the verdict-differential effect.  A stable $g$ does not imply a stable $h$.

\subsection{Measuring variation across eight interfaces}

Because the design is a complete $2^3$ factorial, the eight values of either object $q\in\{g,h\}$ admit an exact Walsh decomposition.  With characters $\chi_S(z)=\prod_{j\in S}z_j$,
\begin{equation}
  \widehat q_S=2^{-3}\sum_z q(z)\chi_S(z),\qquad
  q(z)=\sum_{S\subseteq\{1,2,3\}}\widehat q_S\chi_S(z).
\end{equation}
The empty coefficient $\widehat q_\varnothing$ is the interface average.  The remaining coefficients measure variation around that average.  We summarize their root-mean-square magnitude as
\begin{align}
 R_{\mathrm{total}}(q)&=\Big(\sum_{S\ne\varnothing}\widehat q_S^2\Big)^{1/2},\\
 R_{\mathrm{first}}(q)&=\Big(\sum_{|S|=1}\widehat q_S^2\Big)^{1/2},\\
 R_{\mathrm{high}}(q)&=\Big(\sum_{|S|\ge2}\widehat q_S^2\Big)^{1/2}.
\end{align}
$R_{\mathrm{first}}$ captures variation attached to individual factors; $R_{\mathrm{high}}$ captures interactions left after a first-order approximation.

\paragraph{Observation 1 (effect of averaging).}
This standard factorial identity clarifies what averaging removes.  Full-orbit averaging retains only $\widehat q_\varnothing$.  Averaging factor $j$ alone removes exactly the coefficients whose subsets contain $j$ and leaves the others.  Parseval also gives
\begin{equation}
  2^{-3}\sum_z\{q(z)-\widehat q_\varnothing\}^2=R_{\mathrm{total}}(q)^2,
\end{equation}
and the mean-squared residual after retaining the mean and three main effects is $R_{\mathrm{high}}(q)^2$.  These identities follow from orthonormality of the Walsh characters.  They make the interface average well defined, but do not make it latent truth or certify that individual interfaces agree.

\subsection{Four distinct retrieval conclusions}

We distinguish four questions:
\begin{enumerate}
\item \textbf{Policy level:} after averaging the tested interfaces, does BGE improve downstream performance relative to BM25, and does that gain differ by semantic verdict?
    \item \textbf{Interface stability:} how much does that estimated effect change across task-equivalent answer interfaces?
    \item \textbf{Policy ordering:} do all interfaces support the same direction of the BM25--BGE comparison?
    \item \textbf{Selection value:} does choosing one global ranker from a conventional interface yield different held-out performance than choosing from the averaged interface?
\end{enumerate}
None logically implies another.  Conflating them can turn a prompt-specific observation into an overly strong retrieval claim.

\section{Prospective Evaluation}

\subsection{Datasets and retrieval policies}

\textbf{RAGuard} contributes 2,119 released test claims with fixed BM25 and BGE-v2-m3 rankings~\cite{zeng2025raguard}.  Thirty-five pools contain fewer than three documents, so both policies expose the top $\min(3,\text{available})$ documents without padding.  \textbf{FEVER} contributes a disjoint 1,000-claim subset with 500 SUPPORTS and 500 REFUTES claims~\cite{thorne2018fever}.  The reference policy is pinned Anserini BM25~\cite{yang2017anserini,robertson2009bm25}; BGE-v2-m3 reranks the same fixed top-200 pool and returns its top three~\cite{chen2024bgem3,baai2024bgereranker}.

\subsection{Answer interfaces and readers}

We score Qwen2.5-3B-Instruct-AWQ, Mistral-7B-Instruct-v0.3, Llama-3.1-8B-Instruct, and Gemma-3-4B-Instruct at pinned revisions~\cite{yang2024qwen25,jiang2023mistral,grattafiori2024llama3,gemmateam2025gemma3}.  RAGuard--Llama, RAGuard--Gemma, and all four FEVER environments form the six-environment confirmatory set.  RAGuard--Qwen and RAGuard--Mistral are fresh eight-interface descriptive bridges for previously studied checkpoints.

Each prompt contains an invariant semantic block and a separate interface block.  Binding, A/B versus X/Y vocabulary, and option order are fully crossed.  We score both allowed labels by teacher-forced complete-sequence log probability and choose the larger semantic likelihood.  Greedy generation with at most four new tokens is diagnostic only.  The estimand is therefore controlled forced-choice scoring behavior, not unconstrained free-form generation.

For example, one FEVER interface defines A as SUPPORTS and B as REFUTES.  A binding swap reverses those meanings, a vocabulary swap uses X/Y instead of A/B, and an order swap exchanges the two displayed definitions.  The claim, evidence, semantic task, retrieval policy, and context remain unchanged.

\subsection{FEVER measurement calibration}

Before downstream evaluation, we calibrated BM25 depth and parameters on 2,000 FEVER training claims using a prespecified grid.  We selected the shallowest configuration satisfying the fixed complete-evidence coverage requirement: $K=200$, $k_1=0.8$, and $b=0.2$.  An initial custom BM25 top-50 pilot had failed this measurement requirement and stopped before downstream policy evaluation; its 1,000 claims are excluded from this study.

We applied this configuration once to a prespecified, disjoint development subset.  Complete-evidence coverage was 91.2\% overall, 92.8\% for SUPPORTS, and 89.6\% for REFUTES.  Downstream reader evaluation began only after this measurement check passed.

\subsection{Statistical analysis}

Claims, rather than interface cells, are the sampling units because every claim is observed under all eight interfaces.  Ten thousand verdict-stratified bootstrap resamples are shared within each dataset across readers, rankers, interfaces, objects, and coefficients.  One 95\% max-$|t|$ family simultaneously covers the 84 nonempty coefficients from six environments, two objects, and seven coefficients per object.  A second family covers the 48 confirmatory $g$ cells.  Simultaneous coefficient rectangles yield lower and upper bounds for each variation norm.  Pointwise intervals for policy levels are descriptive and use the same shared resamples~\cite{efron1994bootstrap}.

We regarded interface instability as broadly replicated only if material variation appeared across datasets and reader families.  The prespecified rule (H1) required at least three of six environments, spanning both datasets and at least two reader families, to have a simultaneous $R_{\mathrm{total}}$ lower bound above 0.015 for $g$ or $h$.  On the balanced-accuracy effect scale, 0.015 is 1.5 percentage points.  If H1 held, a higher-order explanation (H2a) required material $R_{\mathrm{high}}$ in at least two environments.  The alternative first-order simplification (H2b) required $R_{\mathrm{high}}$ upper bounds below 0.015 for both objects in at least five environments, jointly replicated first-order signs in at least four, and held-out maximum-error upper bounds below 0.03 in all six.  These thresholds were fixed before downstream evaluation.  Not meeting H1 does not imply equivalence unless the separate simplification conditions also hold.

\subsection{Ranker selection and held-out value}

For each of eight dataset--reader environments, 200 fixed verdict-stratified 40/60 validation/test splits compare two selectors.  Both choose one global ranker: one uses validation balanced accuracy at the canonical interface, and the other first averages semantic log odds over all eight interfaces.  Ties go to BM25.  Validation labels estimate global ranker quality; neither selector conditions on the unknown semantic state of an individual test claim.  Test value is balanced accuracy under interface-averaged decisions.  We define $\Delta V=V_{\mathrm{orbit}}-V_{\mathrm{canonical}}$, so positive values favor the orbit-based selector.  An operational result requires at least 10\% selector disagreement and a 95\% interval for $\Delta V$ materially outside $[-0.005,0.005]$, a half-percentage-point held-out balanced-accuracy threshold fixed before downstream evaluation.

\section{Results}

\subsection{Measurement and execution validity}

Forced-choice likelihood scores were finite for every evaluated condition.  Diagnostic generation invalidity ranges from 0.0\% to 79.3\%, while generation-to-score disagreement ranges from 0.07\% to 79.4\%; both maxima occur for Mistral.  We therefore restrict all conclusions to forced-choice evaluation: we neither repair scores with generations nor make claims about unconstrained generation.

\begin{table*}[t]
\centering
\caption{Confirmatory policy levels and interface variation.  Panel A reports the canonical-cell effect $g_c=g(+,+,+)$ and interface means $\bar g,\bar h$ with descriptive 95\% pointwise intervals; these level estimates are not multiplicity-adjusted and do not enter the prespecified cross-setting decision.  Panel B reports simultaneous intervals for the 84-coefficient family; its final column asks whether the simultaneous lower bound on either total-variation norm exceeds .015.}
\label{tab:spectral}
\begingroup
\footnotesize
\setlength{\tabcolsep}{2.5pt}
\renewcommand{\arraystretch}{1.08}
\textbf{A. Policy levels}\par\smallskip
\begin{tabular}{llccc}
\toprule
Dataset & Reader & $g_c$ [pointwise] & $\bar g$ [pointwise] & $\bar h$ [pointwise] \\
\midrule
RAGuard & Llama & -.0087 [-.0275,.0098] & -.0019 [-.0123,.0084] & .0157 [.0052,.0261] \\
RAGuard & Gemma & .0018 [-.0067,.0103] & .0014 [-.0073,.0099] & .0090 [.0003,.0177] \\
FEVER & Qwen & .0450 [.0150,.0750] & .0399 [.0177,.0620] & .0919 [.0698,.1137] \\
FEVER & Mistral & .0060 [-.0200,.0310] & .0169 [-.0025,.0356] & .0414 [.0227,.0608] \\
FEVER & Llama & .0230 [.0050,.0420] & .0254 [.0116,.0392] & -.0031 [-.0168,.0107] \\
FEVER & Gemma & -.0030 [-.0300,.0240] & .0016 [-.0214,.0245] & .0384 [.0160,.0613] \\
\bottomrule
\end{tabular}
\par\medskip
\textbf{B. Interface variation}\par\smallskip
\begin{tabular}{llccccc}
\toprule
Dataset & Reader & $R_{\rm total}(g)$ [sim.] & $R_{\rm high}(g)$ [sim.] & $R_{\rm total}(h)$ [sim.] & $R_{\rm high}(h)$ [sim.] & LB$>.015$? \\
\midrule
RAGuard & Llama & .005 [.000,.029] & .004 [.000,.023] & .010 [.000,.034] & .007 [.000,.026] & No \\
RAGuard & Gemma & .003 [.000,.015] & .002 [.000,.011] & .002 [.000,.014] & .002 [.000,.011] & No \\
FEVER & Qwen & .025 [.001,.057] & .025 [.001,.052] & .015 [.000,.049] & .014 [.000,.042] & No \\
FEVER & Mistral & .016 [.000,.045] & .003 [.000,.020] & .019 [.000,.047] & .011 [.000,.027] & No \\
FEVER & Llama & .006 [.000,.031] & .006 [.000,.026] & .018 [.000,.043] & .016 [.000,.035] & No \\
FEVER & Gemma & .004 [.000,.029] & .004 [.000,.025] & .012 [.000,.036] & .011 [.000,.033] & No \\
\bottomrule
\end{tabular}
\endgroup
\end{table*}

\begin{table}[t]
\centering
\caption{Descriptive full-interface bridge results for the two previously studied RAGuard readers.  These rows do not enter the six-environment replication rule or its simultaneous family.}
\label{tab:bridge}
\small
\begin{tabular}{lccccc}
\toprule
Reader & $g_c$ & $\bar g$ & $\bar h$ & $R_{\rm total}(g)$ & $R_{\rm total}(h)$ \\
\midrule
Qwen & .0014 & .0018 & .0054 & .0008 & .0038 \\
Mistral & .0069 & .0125 & .0168 & .0044 & .0084 \\
\bottomrule
\end{tabular}
\end{table}

\subsection{Average retrieval-policy effects}

Panel A of Table~\ref{tab:spectral} separates policy level from interface variation.  The descriptive orbit-average BGE-minus-BM25 estimate is .0399 on FEVER--Qwen, with pointwise interval [.0177,.0620], and .0254 on FEVER--Llama, [.0116,.0392].  The other four confirmatory $\bar g$ intervals include zero.  These level estimates show where BGE appears useful overall; they do not establish that the conclusion is stable across interfaces.

Several interface-averaged verdict-differential estimates also have pointwise intervals excluding zero, including FEVER--Qwen ($\bar h=.0919$), FEVER--Mistral ($.0414$), and FEVER--Gemma ($.0384$).  Because these level tests were not part of the prespecified simultaneous decision family, we report them descriptively and do not promote them to confirmatory findings.

\subsection{Does interface instability replicate?}

It does not meet the prespecified cross-setting standard.  None of the six confirmatory environments has a simultaneous $R_{\mathrm{total}}$ lower bound above 0.015 for either $g$ or $h$ (0/6).  FEVER--Qwen has a nonzero higher-order $g$ coefficient, but its variation-norm lower bound is .001, below the materiality threshold.  RAGuard--Gemma is the only environment whose simultaneous upper bounds put higher-order structure below .015 for both objects.

That last result matters for interpretation: only 1/6 environments satisfies the higher-order equivalence component.  Five settings remain too uncertain to satisfy this prespecified higher-order condition.  The prospective data therefore do not support broad replicated material instability, but they also do not establish universal invariance.

\begin{table}[t]
\centering
\caption{Summary of the prespecified hypotheses.  Their formal conditions are described in Section~4 and supplied with the artifacts.}
\label{tab:outcome}
\small
\begin{tabular}{p{0.45\columnwidth}p{0.43\columnwidth}}
\toprule
Question & Observed result \\
\midrule
Broad replicated instability (H1) & Not supported (0/6) \\
Material higher-order instability (H2a) & Not supported (0/6) \\
First-order shortcut (H2b) & Not certified (1/6 equivalence; 0/6 signs; 0/6 error) \\
Certified policy-sign reversal & None \\
Material selector/value effect (H3) & None (0/8) \\
\bottomrule
\end{tabular}
\end{table}

\subsection{Does the policy ordering change?}

No confirmatory environment has simultaneous evidence for either a uniform BM25--BGE direction or a sign reversal.  The 48-cell family classifies all six environments as uncertain.  Some point estimates differ in sign, but we call a reversal only when opposite-direction simultaneous intervals both exclude zero within the same environment.  No environment meets that requirement.

\subsection{Does interface choice affect selection value?}

Interface choice can change which global ranker a validation split selects.  Table~\ref{tab:selection} shows descriptive disagreement estimates as high as .335 for RAGuard--Qwen.  However, none of the eight environments combines at least 10\% disagreement with a material held-out value difference whose 95\% interval excludes $[-.005,.005]$ (0/8).  Selector disagreement alone is therefore not evidence that one selection rule produces better held-out retrieval decisions.

\newpage
\begin{samepage}
\subsection{From the motivating anomaly to prospective replication}

The original RAGuard observation motivated the full-interface audit.  The fresh Qwen and Mistral bridge estimates in Table~\ref{tab:bridge} are descriptive: their total variation is small in the completed eight-interface analysis, but they cannot establish the new cross-setting claim.  Only the prospectively designated environments determine the replication result.  These are RAGuard--Llama, RAGuard--Gemma, and the four FEVER environments.
\end{samepage}

\begin{table}[t]
\centering
\caption{Global ranker-selection disagreement and held-out value difference.  $\Delta V$ is orbit-selector minus canonical-selector balanced accuracy under interface-averaged test decisions.}
\label{tab:selection}
\scriptsize
\resizebox{\columnwidth}{!}{%
\begin{tabular}{llccl}
\toprule
Data & Reader & Disagree. [95\% CI] & $\Delta V$ [95\% CI] & Material? \\
\midrule
RAGuard & Qwen & .335 [.080,.635] & -.0006 [-.0012,.0021] & No \\
RAGuard & Mistral & .185 [.025,.465] & -.0011 [-.0028,.0054] & No \\
RAGuard & Llama & .260 [.045,.600] & -.0019 [-.0039,.0072] & No \\
RAGuard & Gemma & .255 [.050,.520] & -.0001 [-.0017,.0037] & No \\
FEVER & Qwen & .005 [.000,.195] & -.0004 [-.0047,.0030] & No \\
FEVER & Mistral & .255 [.025,.610] & .0049 [-.0034,.0172] & No \\
FEVER & Llama & .030 [.000,.300] & .0002 [-.0039,.0043] & No \\
FEVER & Gemma & .230 [.030,.475] & .0006 [-.0041,.0104] & No \\
\bottomrule
\end{tabular}
}
\end{table}

\section{Discussion}

\paragraph{What did we learn?}
The motivating interface anomaly did not generalize into replicated material instability of the BM25--BGE conclusion under the prospective design.  A large isolated cell was therefore not enough to predict a cross-dataset, cross-family phenomenon.  The experiment also shows why policy level and stability must be displayed together: BGE has positive descriptive average gains in two FEVER environments even though no environment establishes material interface variation.

\paragraph{What does this not mean?}
It does not mean that answer interfaces are universally interchangeable.  Five of six confirmatory environments remain too uncertain for the higher-order equivalence component, and diagnostic generations can disagree sharply with forced-choice scores.  Nor does interface variation reveal a particular token prior, semantic prior, or reader mechanism.  The conclusion is restricted to the enumerated labels, two datasets, four readers, one context depth, and this BM25--BGE comparison.

\paragraph{What should retrieval researchers report?}
A single prompt, significant interface cell, or raw selector disagreement should not be promoted into a general claim that a reranker helps or harms.  When the downstream answer interface is arbitrary and stability is unverified, reporting an interface-averaged policy effect together with explicit bounds on interface variation is conservative.  This recommendation is conditional: the evidence does not imply that every RAG evaluation must enumerate these exact eight interfaces.

\paragraph{Simpler reporting choices provide different protections.}
A canonical-interface evaluation is inexpensive, but it provides no evidence that the resulting policy conclusion is stable.  Swapping only the label binding is a useful common counterbalance: it removes variation attached to that factor, while leaving vocabulary, order, and their interactions unresolved.  Uniform averaging over the complete enumerated orbit is the strongest simple summary considered here.  It returns the orbit-average policy effect by construction, but it does not certify small variation, preserved ordering, or downstream selection value; those are separate inferential questions.  Finally, a reduced first-order approximation is cheaper than the complete factorial analysis.  Its prespecified low-order adequacy criterion held in only one of six confirmatory environments, so our results do not justify replacing the full audit with that approximation.  These summaries are complements rather than interchangeable substitutes: each answers a progressively stronger question about a retrieval-policy conclusion.

\section{Reproducibility}

The supplement contains the prespecified protocol, claim-level predictions, analysis code, prompts, and reproducibility metadata.  Dataset and model versions are pinned.  Experiments used one NVIDIA GeForce RTX 5090 for 5.80 successful GPU-hours, with peak recorded VRAM of 20,707 MiB; detailed runtime, precision, and device metadata are provided in the supplement.

\section{Limitations}

The audit covers a finite interface with two neutral vocabularies, not arbitrary wording.  It uses forced-choice likelihood rather than free generation; diagnostic generation-to-score disagreement reaches 79.4\%.  The evaluation includes two fact-checking datasets, four small open readers, BM25 and BGE-v2-m3, one context depth, and fixed candidate pools rather than live Web retrieval.  Complete-evidence coverage depends on dataset annotations, balanced accuracy does not encode application-specific costs, and the selector analysis is an offline split experiment.  These boundaries preclude claims about other retrieval treatments, free-form reader behavior, or deployed systems under distribution shift.

\section{Conclusion}

We asked whether arbitrary answer-interface choices change conclusions about a fixed BM25--BGE retrieval comparison.  Across RAGuard and FEVER, four readers, and all eight combinations of label binding, vocabulary, and option order, none of the six confirmatory environments certified material variation, no policy-sign reversal was certified, and none of eight selector tests established a material held-out value difference.  The motivating anomaly therefore did not generalize as a broad material phenomenon under the tested conditions.  This is not proof of universal invariance.  Five settings remain too uncertain to satisfy the prespecified higher-order equivalence condition, and the estimand concerns forced-choice scores rather than free generation.  The practical lesson is to keep average policy gain, interface stability, policy ordering, and selection value separate.  When stability is unresolved, an interface-averaged estimate reported with explicit variation bounds is a conservative basis for a retrieval claim.

\section{Ethical Considerations}

The study uses released fact-checking datasets and public model checkpoints, with no human-subject data collection or live intervention.  Its main ethical relevance is epistemic: attributing a measurement artifact to a retrieval policy can support the wrong system decision.  Dataset errors, model biases, and misleading evidence remain possible, and the study makes no deployment claim.  Interface averaging is not evidence that an answer is truthful or unbiased.

\bibliographystyle{ACM-Reference-Format}
\bibliography{bibliography}

@inproceedings{lewis2020rag,
  author = {Patrick Lewis and Ethan Perez and Aleksandra Piktus and Fabio Petroni and Vladimir Karpukhin and Naman Goyal and Heinrich K{\"u}ttler and Mike Lewis and Wen-tau Yih and Tim Rockt{\"a}schel and Sebastian Riedel and Douwe Kiela},
  title = {Retrieval-Augmented Generation for Knowledge-Intensive {NLP} Tasks},
  booktitle = {Advances in Neural Information Processing Systems},
  year = {2020}
}

@inproceedings{thorne2018fever,
  author = {James Thorne and Andreas Vlachos and Christos Christodoulopoulos and Arpit Mittal},
  title = {{FEVER}: a Large-scale Dataset for Fact Extraction and {VER}ification},
  booktitle = {Proceedings of NAACL-HLT},
  pages = {809--819},
  year = {2018},
  doi = {10.18653/v1/N18-1074}
}

@inproceedings{zeng2025raguard,
  author = {Linda Zeng and Rithwik Gupta and Divij Motwani and Yi Zhang and Diji Yang},
  title = {Worse than Zero-shot? A Fact-Checking Dataset for Evaluating the Robustness of {RAG} Against Misleading Retrievals},
  booktitle = {Advances in Neural Information Processing Systems},
  year = {2025},
  url = {https://arxiv.org/abs/2502.16101}
}

@inproceedings{yang2017anserini,
  author = {Peilin Yang and Hui Fang and Jimmy Lin},
  title = {{Anserini}: Enabling the Use of {Lucene} for Information Retrieval Research},
  booktitle = {Proceedings of SIGIR},
  pages = {1253--1256},
  year = {2017},
  doi = {10.1145/3077136.3080721}
}

@article{robertson2009bm25,
  author = {Stephen Robertson and Hugo Zaragoza},
  title = {The Probabilistic Relevance Framework: {BM25} and Beyond},
  journal = {Foundations and Trends in Information Retrieval},
  volume = {4},
  number = {1--2},
  pages = {1--174},
  year = {2009},
  doi = {10.1561/1500000019}
}

@inproceedings{chen2024bgem3,
  author = {Jianlyu Chen and Shitao Xiao and Peitian Zhang and Kun Luo and Defu Lian and Zheng Liu},
  title = {{M3-Embedding}: Multi-Linguality, Multi-Functionality, Multi-Granularity Text Embeddings Through Self-Knowledge Distillation},
  booktitle = {Findings of the Association for Computational Linguistics: ACL 2024},
  pages = {2318--2335},
  year = {2024},
  doi = {10.18653/v1/2024.findings-acl.137}
}

@misc{baai2024bgereranker,
  author = {{Beijing Academy of Artificial Intelligence}},
  title = {{bge-reranker-v2-m3} Model Card},
  year = {2024},
  howpublished = {Hugging Face model repository},
  url = {https://huggingface.co/BAAI/bge-reranker-v2-m3},
  note = {Accessed 2026-08-10}
}

@article{yang2024qwen25,
  author = {{Qwen Team}},
  title = {{Qwen2.5} Technical Report},
  journal = {arXiv preprint arXiv:2412.15115},
  year = {2024}
}

@article{jiang2023mistral,
  author = {Albert Q. Jiang and Alexandre Sablayrolles and Arthur Mensch and Chris Bamford and Devendra Singh Chaplot and Diego de las Casas and Florian Bressand and Gianna Lengyel and Guillaume Lample and Lucile Saulnier and L{\'e}lio Renard Lavaud and Marie-Anne Lachaux and Pierre Stock and Teven Le Scao and Thibaut Lavril and Thomas Wang and Timoth{\'e}e Lacroix and William El Sayed},
  title = {Mistral 7B},
  journal = {arXiv preprint arXiv:2310.06825},
  year = {2023}
}

@article{grattafiori2024llama3,
  author = {Aaron Grattafiori and others},
  title = {The Llama 3 Herd of Models},
  journal = {arXiv preprint arXiv:2407.21783},
  year = {2024}
}

@article{gemmateam2025gemma3,
  author = {{Gemma Team}},
  title = {Gemma 3 Technical Report},
  journal = {arXiv preprint arXiv:2503.19786},
  year = {2025}
}

@inproceedings{zheng2024pride,
  author = {Chujie Zheng and Hao Zhou and Fandong Meng and Jie Zhou and Minlie Huang},
  title = {Large Language Models Are Not Robust Multiple Choice Selectors},
  booktitle = {International Conference on Learning Representations},
  year = {2024},
  url = {https://arxiv.org/abs/2309.03882}
}

@inproceedings{liusie2024general,
  title = {Teacher-Student Training for Debiasing: General Permutation Debiasing for Large Language Models},
  author = {Adian Liusie and Yassir Fathullah and Mark J. F. Gales},
  booktitle = {Findings of the Association for Computational Linguistics: ACL 2024},
  year = {2024}
}

@inproceedings{wang2024firsttoken,
  author = {Xinpeng Wang and Bolei Ma and Chengzhi Hu and Leon Weber-Genzel and Paul R{\"o}ttger and Frauke Kreuter and Dirk Hovy and Barbara Plank},
  title = {{``My Answer is C''}: First-Token Probabilities Do Not Match Text Answers in Instruction-Tuned Language Models},
  booktitle = {Findings of the Association for Computational Linguistics: ACL 2024},
  year = {2024}
}

@inproceedings{wang2024looktext,
  author = {Xinpeng Wang and Chengzhi Hu and Bolei Ma and Paul R{\"o}ttger and Barbara Plank},
  title = {Look at the Text: Instruction-Tuned Language Models are More Robust Multiple Choice Selectors than You Think},
  booktitle = {Conference on Language Modeling},
  year = {2024}
}

@inproceedings{cho2025hidden,
  author = {Hakaze Cho and Yoshihiro Sakai and Mariko Kato and Kenshiro Tanaka and Akira Ishii and Naoya Inoue},
  title = {Token-based Decision Criteria Are Suboptimal in In-context Learning},
  booktitle = {Proceedings of NAACL-HLT},
  year = {2025},
  doi = {10.18653/v1/2025.naacl-long.278}
}

@article{huang2026yesno,
  author = {Haonan Huang},
  title = {The yes-no bias of large language models reflects answer order and wording, not shifts in moral judgment},
  journal = {arXiv preprint arXiv:2607.05552},
  year = {2026}
}

@inproceedings{huang2026pcfjudge,
  title = {Permutation-Consensus Listwise Judging for Robust Factuality Evaluation},
  author = {Tianyi Huang and Nathan Huang and Justin Tang and Wenqian Chen and Elsa Fan},
  booktitle = {Proceedings of the Fifth Workshop on Generation, Evaluation and Metrics},
  pages = {595--603},
  year = {2026},
  doi = {10.18653/v1/2026.gem-main.58}
}

@article{liao2026evidence,
  author = {Junchi Liao and Jiawen Deng and Fuji Ren},
  title = {Evidence Interfaces Shape How Retrieval-Augmented Readers Use Support},
  journal = {arXiv preprint arXiv:2607.17108},
  year = {2026}
}

@article{thomas2026forminv,
  author = {Nishal Thomas and Noel Thomas},
  title = {{FormInv}: A Measurement Protocol for Semantic Invariance in Mathematical Reasoning Benchmarks},
  journal = {arXiv preprint arXiv:2605.29001},
  year = {2026},
  url = {https://arxiv.org/abs/2605.29001}
}

@inproceedings{polo2024prompteval,
  author = {Felipe Maia Polo and Ronald Xu and Lucas Weber and M{\'i}rian Silva and Onkar Bhardwaj and Leshem Choshen and Allysson Flavio Melo de Oliveira and Yuekai Sun and Mikhail Yurochkin},
  title = {Efficient Multi-Prompt Evaluation of {LLM}s},
  booktitle = {Advances in Neural Information Processing Systems},
  year = {2024},
  url = {https://arxiv.org/abs/2405.17202}
}

@inproceedings{hsia2025ragged,
  author = {Jennifer Hsia and Afreen Shaikh and Zhiruo Wang and Graham Neubig},
  title = {{RAGGED}: Towards Informed Design of Scalable and Stable {RAG} Systems},
  booktitle = {Proceedings of the International Conference on Machine Learning},
  year = {2025},
  url = {https://arxiv.org/abs/2403.09040}
}

@inproceedings{sun2025promptrankers,
  author = {Shuoqi Sun and Shengyao Zhuang and Shuai Wang and Guido Zuccon},
  title = {An Investigation of Prompt Variations for Zero-shot {LLM}-based Rankers},
  booktitle = {Proceedings of the 47th European Conference on Information Retrieval},
  year = {2025},
  url = {https://arxiv.org/abs/2406.14117}
}

@book{efron1994bootstrap,
  author = {Bradley Efron and Robert J. Tibshirani},
  title = {An Introduction to the Bootstrap},
  publisher = {CRC Press},
  year = {1994}
}

\end{document}